# Current-sensitive Hall effect in a chiral-orbital-current state

Yu Zhang[1], Yifei Ni[1], Pedro Schlottmann[2], Rahul Nandkishore[1,3] Lance E. DeLong[4], and Gang Cao[1,5*]

[1]*Department of Physics, University of Colorado at Boulder, Boulder, CO 80309, USA*

[2]*Department of Physics, Florida State University, Tallahassee, FL 32306, USA*

[3]*Center for Theory of Quantum Matter, University of Colorado at Boulder, Boulder, CO 80309, USA*

[4]*Department of Physics and Astronomy, University of Kentucky, Lexington, KY 40506, USA*

[5]*Center for Experiments on Quantum Materials, University of Colorado at Boulder, Boulder, CO 80309, USA*

Chiral orbital currents (COC) underpin a novel colossal magnetoresistance (CMR) in ferrimagnetic $Mn_3Si_2Te_6$ [1]. Here we report the Hall effect in the COC state which exhibits the following unprecedented features: (1) A sharp, current-sensitive peak in the magnetic field dependence of the Hall resistivity; (2) An unusually large Hall angle reaching up to 0.15 (comparable to the highest values yet reported); and (3) A current-sensitive scaling relation between the Hall conductivity $\sigma_{xy}$ and the longitudinal conductivity $\sigma_{xx}$, namely, $\sigma_{xy} \propto \sigma_{xx}^{\alpha}$ with $\alpha$ ranging between 3 and 5, which is both sensitive to external current and exceptionally large compared to $\alpha \leq 2$ typical of most solids. These anomalies point to a giant, current-sensitive Hall effect that is unique to the COC state. We argue that a magnetic field induced by the fully developed COC combines with the applied magnetic field to exert the greatly enhanced transverse force on charge carriers, which dictates the novel Hall responses. The COC Hall effect is unique, as it is generated and controlled via the *interaction between intrinsic COC and applied external currents*, which leads to novel transport phenomena of fundamental and technological significance and requires new physics for explanation.

*Corresponding author. Email: gang.cao@colorado.edu

Our recent study [1] revealed chiral orbital currents (COC) in a colossal magnetoresistance (CMR) material, ferrimagnetic $Mn_3Si_2Te_6$ (**Figs. 1a-1c**). The COC circulate along the edges of $MnTe_6$ octahedra to underpin an astonishing $10^7$-CMR that occurs without a net magnetic polarization along the magnetic hard axis (**Fig. 1f**) [1-3]. The hallmark of this COC state, $\Psi_C$, is its unusually high sensitivity to small external currents [1].

Ferrimagnetic $Mn_3Si_2Te_6$ with trigonal symmetry (*P-31c*) [4-7] orders at a transition temperature $T_C$ = 78 K, with the magnetic easy axis along the *a* axis and the magnetic hard axis along the *c* axis (**Fig. 1f**) [1-10]. There are two inequivalent Mn1 and Mn2 sites in the unit cell. The $MnTe_6$ octahedra form a honeycomb sublattice of Mn1 ions in the *ab* plane (**Fig. 1b**), whereas the $MnTe_6$ octahedra form a triangular sublattice of Mn2 ions sandwiched between the honeycomb layers (**Figs. 1a-1b**) [4].

A recent neutron diffraction study reveals a noncollinear magnetic structure with the magnetic space group *C2'/c'*, where the Mn spins lie predominantly within the *ab* plane but tilt toward the *c* axis by ~ 10° in ambient conditions (**Fig. 1c**) [8], which simultaneously breaks mirror and time reversal symmetries [8, 9]. The COC state arises below $T_C$ with the orbital currents predominantly within the *ab* plane (**Figs. 1a-1b**) [1], which generate orbital moments $M_{COC}$ primarily oriented along the *c* axis (**Fig. 1a**). (Although the orbital moments could interact with each other at H = 0 this causes no additional order, likely due to strong fluctuations that are evident in previous studies [2, 8].) The $M_{COC}$ is estimated to be on the order of 0.1 $\mu_B$ [1], and is coupled with the Mn spins, which yields an unusual spin-orbit effect that produces a large anisotropy field of 13 T (note that the *orbital angular momentum* is zero for the $Mn^{2+}$ ($3d^5$) ion with a half-filled 3*d* orbital) [1, 2, 4]. In the absence of a magnetic field **H || *c*** axis, the net circulation of the COC is zero since it can circulate both clockwise and counterclockwise (**Fig. 1d)**. This results in

disordered circulation domains that cause strong scattering and high resistance. However, application of **H || *c*** axis favors only one direction of circulation (i.e., either clockwise or counterclockwise) and expands its domains, concurrently reducing and eventually suppressing other domains with the opposite direction of circulation (**Fig. 1e**). The increased size of the preferred COC domains (and concurrent decrease in domain wall volume) leads to a sharp reduction in electron scattering, and thus the $10^7$-CMR (**Fig. 1f**) [1]. The COC as intrinsic currents are unusually susceptible to externally applied currents I that disrupt and eventually "melt" the COC state $\Psi_C$, resulting in a first-order transition to a trivial state $\Psi_T$ when I exceeds a critical threshold [1].

The interaction between COC and I presents new, intriguing physics that needs to be understood. We have chosen to apply the Hall effect as a fundamental, powerful probe of this interaction. The *ordinary* Hall effect in solids induces a transverse voltage drop proportional to H resulting from the Lorentz force on electrons. The *anomalous* Hall effect (AHE) usually arises from a ferromagnetic state with broken time-reversal symmetry [11]. In particular, electrons in the presence of a scalar spin chirality $S_i \bullet (S_j \times S_k)$ gain an anomalous transverse velocity [12] or Berry phase curvature [13], which in turn adds to the Hall effect as an intrinsic contribution independent of scattering. The Hall resistivity $\rho_{xy}$ is thus anticipated to be proportional to its magnetization M [14-17]. In other magnets such as the helical magnet MnSi, $\rho_{xy}$ exhibits an unusual stepwise field profile that is attributed to an effective magnetic field due to chiral spin textures [18]. The Berry phase is also linked to a *topological* Hall effect (THE) observed in certain topological semimetals with a strong spin-orbit interaction (SOI) [e.g., 17-21]. Recent studies indicate that an intrinsic AHE can occur in a noncollinear antiferromagnet with strong SOI so long as mirror and time reversal symmetries both are broken [22, 23].

Here we report a *strongly-current-sensitive* Hall effect in ferrimagnetic $Mn_3Si_2Te_6$ that exhibits the following novel behaviors: (1) A distinct, sharp peak in the field dependence of $\rho_{xy}$ is a sensitive function of I (**Fig. 1g**). (2) An unusually large Hall angle (given by the ratio of the Hall conductivity $\sigma_{xy}$ to the longitudinal conductivity $\sigma_{xx}$) reaches up to 0.15 which is comparable to values reported in magnets having a giant Hall effect [20, 22-25]. (3) A scaling relation $\sigma_{xy} \propto \sigma_{xx}^{\alpha}$ is obeyed with $\alpha$-values ranging between 3 and 5, which are *unprecedentedly large* compared to $\alpha \leq 2$ typical of most solids [17], and *sensitively depend on I.*

An exceptionally large $\alpha$ indicates that in the COC state, $\sigma_{xy}$ rises with H much faster than $\sigma_{xx}$. We argue that the *c*-axis orbital moments $M_{COC}$ induced by the COC produce a real-space magnetic field $\mathbf{b_c}$ that adds to an applied field $\mathbf{H} \parallel \boldsymbol{c}$ axis, i.e., $\mathbf{H} + \mathbf{b_c}$; as such the charge carriers gain an additional transverse velocity that generates the giant, current-sensitive Hall effect. This Hall effect shows no simple correlation with the magnetization M (**Figs. 1f-1g**), as predicted by the Karplus-Luttinger theory [12], nor does it behave as observed or expected in other materials [17]. We will refer to this unique Hall effect as the *COC Hall effect (COCHE)*.

Experimental details, including measurement techniques and processes, and additional data are described in *Methods and Extended Data Figs.1-4*. Note that a self-heating effect is completely ruled out, which is discussed in [1] and *Methods and Extended Data Figs.2-3.*

We first focus on $\rho_{xy}$ as a function of $\mathbf{H} \parallel \boldsymbol{c}$ axis at T = 30 K as an example (**Fig. 2**). Note that $\rho_{xy}$ exhibits a sharp peak at a critical field $H_C$ that marks an onset of the COC state $\Psi_C$. The peak is then followed by a rapid decrease of $\rho_{xy}$ by up to two orders of magnitude (**Figs. 2a-2e**). (Note that I = 1 mA corresponds to a modest current density J $\approx$ 1 A/cm$^2$ in the samples measured; for clarity we use I in the discussion.) The peak shifts to higher fields with increasing I, revealing a sharp switching at I = 3 mA and 4.5 mA (**Figs. 2c-2d**) before evolving into a broader peak at I =

5 mA (**Fig. 2e**), which signals a vanishing $\Psi_C$ and an emerging trivial state $\Psi_T$. This behavior indicates that the COC weaken as I is increased, and thus stronger **H || *c*** axis are required to offset the disruption of the COC state caused by I (see discussion below). The correlation between I and $H_C$ for T =30 K is illustrated in **Fig. 2f**. The peak at $H_C$ (an indicator for $\Psi_C$) can persist up to $T_C$ = 78 K so long as I is small (e.g., 1 mA; see **Figs. 2g-2i** for selected temperatures T). (Note that a strong hysteresis in $\rho_{xy}$ and $\rho_{xx}$ is seen between H ramping up and down [see *Extended Data Fig.1*], as well as in previous studies [1, 2], consistent with the presence of COC domains discussed above.) However, larger I exceeding a certain threshold value $I_C$ can readily suppress the COC and recover $\Psi_T$ even well below $T_C$; this happens, for example, at 30 K when I ≥ 5 mA. In these cases, the field dependence of $\rho_{xy}$ exhibits a behavior similar to that at 100 K and I = 1 mA, which is a benchmark for $\Psi_T$ that is signaled by only a broad or suppressed peak in the field dependence of $\rho_{xy}$ (**Fig. 2j**).

In $\Psi_C$ at H > $H_C$, $\rho_{xy}$ increases slowly and linearly with H (see left scale in **Fig. 2k**). However, as $\Psi_C$ weakens and eventually transitions to $\Psi_T$ with increasing I, the slope of $\rho_{xy}$ evolves from positive to negative (right scale in **Fig. 2k**). The carrier density n (~ H/$\rho_{xy}$) is thus a strong function of I (see **Figs. 2l - 2o** for representative T). At low T, n(I) decreases with increasing I monotonically and with a sign change at I > $I_C$ (**Fig. 2l**) that marks a change of state. At T = 50 K and 70 K, n(I) remains positive in $\Psi_C$ and becomes negative in $\Psi_T$ (**Fig. 2m-2n**). At T = 100 K > $T_C$, n(I) stays negative (**Fig. 2o**). These observations allow us to conclude that the charge carriers are primarily holes in $\Psi_C$ and electrons in $\Psi_T$ between $T_C$ and 120 K [*Methods*].

In sharp contrast to $\rho_{xy}$ for **H || *c*** axis, $\rho_{xy}$ for **H || *b*** axis exhibits a field dependence that is more consistent with an ordinary Hall effect (e.g., **Fig. 1g** and *Extended Data Fig.3*) with no discernible evidence of an AHE, despite the fact that the *a* axis is the magnetic easy axis, and the *a*-axis magnetization $M_a$ is fully saturated, reaching 1.6 $\mu_B$/Mn at $\mu_o$H < 0.1 T (**Fig. 1c**) [1, 2]. The

corresponding n(I) is on the order of $10^{23}/m^3$, comparable to that for $\Psi_T$ with **H || *c*** axis (**Fig. 2o**). The contrasting Hall responses observed for **H || *c*** axis and **H || *a*** axis signal a highly anisotropic band structure [10], and more generally the novelty of the COCHE.

Moreover, for **H || *c*** axis and T < $T_C$, the Hall angle, defined as the ratio of the Hall conductivity $\sigma_{xy}\left(=\frac{\rho_{xy}}{\rho_{xx}^2+\rho_{xy}^2}\right)$ to the longitudinal conductivity $\sigma_{xx}\left(=\frac{\rho_{xx}}{\rho_{xx}^2+\rho_{xy}^2}\right)$, or $\sigma_{xy}/\sigma_{xx}$, rises drastically when H enters a certain crossover region. As shown in **Figs. 3a-3b**, $\sigma_{xy}/\sigma_{xx}$ initially increases slowly with H and remains smaller than 0.01 below 9 T. However, above the crossover region marked by the gray band in **Figs. 3a-3b**, $\sigma_{xy}/\sigma_{xx}$ rapidly rises, reaching up to 0.15, which is a value comparable to the largest reported in magnets having a giant Hall effect [21-25]. It indicates that $\sigma_{xy}$ increases much faster than $\sigma_{xx}$ with increasing H when the COC state is fully developed in higher fields (> 6 T); note that the onset of the CMR occurs at $\mu_o H_{||c} \approx 3$ T (**Fig. 1f**) [1, 2]. Above $T_C$ where $\Psi_T$ prevails, $\sigma_{xy}/\sigma_{xx}$ increases with H slowly and linearly, similar to the behavior seen below the crossover region (**Figs. 3a-3b**). Remarkably, near the crossover region, $\sigma_{xy}/\sigma_{xx}$ exhibits a brief, yet prominent inverted peak at I = 2 mA when T = 70 K (**Fig. 3a**) and at I = 3 mA when T = 50 K (**Fig. 3b)**. This peak persistently occurs whenever the system approaches the vicinity of the transition between $\Psi_C$ and $\Psi_T$ and is discussed further below.

We now examine the behavior of the scaling relation $\sigma_{xy} \propto \sigma_{xx}^{\alpha}$ for a few representative I and T. Below $T_C$, $\sigma_{xy}$ scales with $\sigma_{xx}$ and generates two different values of the exponent, namely $\alpha_{HF}$ (obtained at higher fields) and $\alpha_{LF}$ (obtained at lower fields), which define two distinct regions corresponding to a fully developed $\Psi_C$ state and $\Psi_T$ or a mixed state of $\Psi_C$ and $\Psi_T$, respectively. A cutoff field that separates $\alpha_{HF}$ and $\alpha_{LF}$ falls in the crossover region marked in **Figs. 3a-3b**. An unanticipated, novel feature of this scaling relation is that *the exponent $\alpha_{HF}$ is both unprecedentedly* *large and sensitive to I.* Specifically, $\alpha_{HF}$ reaches 5 at I = 1 mA, as shown in **Fig. 3c** where the data above the cutoff field (9.5 T) perfectly trace the scaling relation $\sigma_{xy} \propto \sigma_{xx}^5$. With increasing I, $\alpha_{HF}$ reduces to 4 and 3 at I = 2 mA and 3 mA, respectively (**Figs. 3e** and **3g**), suggesting that $\Psi_C$ gets weakened with increasing I. Below the cutoff field (shaded regions in **Figs. 3c, 3e, 3g**), $\sigma_{xy}$ conforms to a scaling relation where $\alpha = \alpha_{LF} = 2.2$, 2.1 and 1.8 at I = 1 mA, 2 mA and 3 mA, respectively (**Figs. 3d, 3f, 3h**), indicating a vanishing $\Psi_C$ and an emerging $\Psi_T$. Applying I = 5 mA suppresses $\Psi_C$ and generates a linear scaling relation $\sigma_{xy} \propto \sigma_{xx}$ at 30 K (**Fig. 3i**). Similarly, a scaling relation with $\alpha_{LF} \leq 2$ is seen at T = 70 K and I = 3 mA [*Extended Data Fig.4*], and T = 100 K and I = 1 mA (**Fig. 3j**). Importantly, $\sigma_{xy}$ and $\sigma_{xx}$ exhibit distinct H dependences at higher H. For example, at I = 2 mA and T = 30 K, $\sigma_{xy}$ rises rapidly and linearly whereas $\sigma_{xx}$ exhibits a tendency of saturation above H > 8.3 T (**Fig. 3k**); therefore, the increase of $\sigma_{xy}$ or $[\sigma_{xy}(14T) - \sigma_{xy}(8.3T)]/\sigma_{xy}(8.3T) = 440\%$ above 8.3 T, but the value for $\sigma_{xx}$ is merely 83%. This implies that an additional driving force strongly affects the transverse current (further discussed below) and explains the unusual scaling relation with the exceptionally large $\alpha$.

A phase diagram generated from the data illustrates that a hallmark of a fully developed $\Psi_C$ is a strongly current-dependent scaling relation $\sigma_{xy} \propto \sigma_{xx}^\alpha$ with an unprecedented range of $3 \leq \alpha = \alpha_{HF} \leq 5$ (**Fig. 3l**). However, when $\Psi_C$ is less robust or suppressed, this unique scaling relation is supplanted by another with $1 \leq \alpha = \alpha_{LF} \leq 2.2$, which is qualitatively similar to the range $1.6 \leq \alpha \leq 2$ commonly observed in insulators and bad metals having strong disorder [17, 25].

We now turn to the Hall effect as a function of T. Both $\rho_{xx}$ and $\rho_{xy}$ peak at T = $T_P$ well above $T_C$ (marked by hollow and solid arrows in **Figs. 4a-4b**). The peak at $T_P$ is due to the broadening of the ferrimagnetic transition [2] by H (in this case, $\mu_o H_{||c}$ = 7 T). Both $T_P$ and $T_C$ progressively shift to lower T with increasing I, but the temperature difference, ΔT, remains essentially unchanged, i.e., $\Delta T = T_P - T_C \approx 32$ K at 7 T. As T decreases, $\rho_{xx}$ drops rapidly over the ΔT interval and reaches its lowest value slightly below $T_C$ (**Fig. 4a**). (Note that at $I \geq 5$ mA, $T_C$ is suppressed, thus $\Psi_T$ emerges.) On the other hand, $\rho_{xy}$ at I = 1 mA drops more than one order of magnitude from $T_P$ to $T_C$ (blue curve in **Fig. 4b**). With increasing I, $\rho_{xy}$ undergoes a rapid sign change from positive to negative to positive again over the span of ΔT. This change results in a sharp inverted peak that progressively amplifies as I increases (**Fig. 4b**). This peak could be a consequence of a transient crystal and/or band structure change driven by I, which could strongly affect the COC. Indeed, it is already established that I can readily alter crystal structures, and therefore, electronic structure in canted magnets with large SOI and nearly degenerate states [26-28]. Nevertheless, the original crystal and/or band structure can be quickly recovered via either further increasing H or decreasing T (see **Figs. 3a-3b** and **Fig. 4b**).

While the microscopic origin of the new COCHE is yet to be established, we argue that the *c*-axis orbital moments $M_{COC}$ induced by the COC play an essential role in this Hall effect. The high sensitivity of the COCHE to small I suggests a very delicate nature of the COC circulating along the edges of $MnTe_6$ octahedra. As already recognized [1], application of **H || *c*** axis expands the *ab*-plane COC domains with one direction of circulation and concurrently shrinks the COC domains with the opposite direction of circulation (**Figs. 1d-1e**). The expanded *ab*-plane COC domains in turn generate stronger *c*-axis $M_{COC}$, which render a magnetic field $\mathbf{b_c}$ aligned along the *c* axis and proportional to H. This induced field $\mathbf{b_c}$ couples with **H || *c*** axis to yield an enhanced effective magnetic field $\mathbf{H} + \mathbf{b_c}$ acting on itinerant holes in the COC state. The itinerant holes are strongly deflected by $\mathbf{H} + \mathbf{b_c}$ and thereby gain a significant, additional transverse velocity, which is reflected by the greatly enhanced Hall current/conductivity schematically illustrated in **Fig. 4c**.

*This scenario qualitatively explains the key observations of this study*. The sharp peak at $H_C$ in the field dependence of $\rho_{xy}$ (**Fig. 2**) signals the emergence of $\mathbf{b_c}$, and the rapid decrease in $\rho_{xy}$ at $H > H_C$ is a consequence of an added Hall current generated by $\mathbf{H} + \mathbf{b_c}$ as $\Psi_C$ is fully developed. When higher I is applied, a stronger $H_C$ is needed (**Fig. 2f**) to offset the damage done to the COC in order to further enhance the COC domains, thus $\mathbf{b_c}$. The unusually large Hall angle and the extraordinarily large $\alpha$ in the current-sensitive $\sigma_{xy} \propto \sigma_{xx}^{\alpha}$ (**Fig. 3**) can be ascribed to the additional transverse velocity of the holes ($\propto \mathbf{H} + \mathbf{b_c}$) that drives an extraordinarily strong increase in $\sigma_{xy}$ with H in the high-field regime where $\mathbf{b_c}$ ($\propto$ H) gets further strengthened; in contrast, $\sigma_{xx}$ in this high-field regime tends to saturate (**Fig. 3k**). This explains that the COC Hall scaling relation with $3 \leq \alpha = \alpha_{HF} \leq 5$ is operative at much higher H only when the COC are fully established. In summary, the current-sensitive *COCHE* is a novel transport phenomenon with great fundamental and technological promise. The novelty of the underlying COC state suggests that other unanticipated types of Hall effect remain to be discovered in materials with a variety of ground states.

## Main figure legends

**Fig.1. Key structural and physical properties. a,** The crystal and magnetic structure of $Mn_3Si_2Te_6$ [1]. The colored circles and vertical arrows indicate the *ab*-plane COC and induced $M_{COC}$, respectively; the faint cylindrical arrows are Mn spins. **b,** The COC circulating in the honeycomb lattice in the *ab* plane [1]. **c**, The canted Mn spins in the *ab* plane [8]. **d-e**, Schematics of the *ab*-plane COC at H = 0 (d) and $H_{||c} > 0$ (e). **f**, The magnetic field dependence of the *a*-axis magnetoresistance ratio $[\rho_{xx}(H)-\rho_{xx}(0)]/\rho_{xx}(0)$ and the magnetization M (dashed lines, right scale) for **H || *c*** axis (blue) and **H || *a*** axis (red) [2]. **g**, The magnetic field dependence of the Hall resistivity $\rho_{xy}$ (H||*c*) (blue) and $\rho_{xy}$ (H||*b*) (red) at T = 30 K and I = 2 mA; the black arrow marks a critical field $H_C$. The insets are configurations for sample (S) measurements of $\rho_{xy}$ (H||*c*) (lower inset) and $\rho_{xy}$ (H||*b*) (upper inset).

**Fig. 2. The Hall effect as functions of magnetic field and external current. a-e,** The magnetic field dependence of $\rho_{xy}$ (H||*c*) at 30 K for selected currents I; the black arrows mark the critical field $H_C$. **f**, The correlation between $H_C$ and I at T = 30 K; **g-j**, The magnetic field dependence of $\rho_{xy}$ (H||*c*) at selected temperatures and currents. **k**, The zoomed-in $\rho_{xy}$ (H||*c*) at T = 30 K (a-e) in a higher-field regime of 9-14 T. **l-o**, The carrier density n estimated from the data of $\rho_{xy}$ (H||*c*) for selected temperatures.

**Fig. 3. The Hall angle and scaling relation as a function external current. a-b,** The magnetic field dependence of the Hall angle $\sigma_{xy}/\sigma_{xx}$ at I = 2 mA (a) and I = 3mA (b) for selected temperatures; the gray band marks a crossover region between $\Psi_C$ and $\Psi_T$. **c-i**, The scaling relation $\sigma_{xy} \propto \sigma_{xx}^{\alpha}$ at T = 30 K for selected currents I. The shaded areas indicate a regime of $\Psi_T$ having a lower value of the exponent $\alpha$; the arrows indicate the cutoff field. **j**, The scaling relation $\sigma_{xy} \propto \sigma_{xx}^{1.8}$ at T = 100 K and I = 1 mA. **k**, $\sigma_{xx}$ (left scale) and $\sigma_{xy}$ (right scale) as a function of **H || *c*** axis at T = 30 K and

$I = 2$ mA; the marked field 8.3 T indicates a crossover field above which $\sigma_{xy}$ rises much faster than $\sigma_{xx}$. **l**, The phase diagram of the exponent $\alpha$ as a function of external current I. Note that $\Psi_C$ features $3 \leq \alpha = \alpha_{HF} \leq 5$ whereas $\Psi_T$ $1 \leq \alpha = \alpha_{LF} \leq 2.2$.

**Fig. 4. The Hall effect and resistivity as functions of temperature and external current.** The temperature dependence of **a**, $\rho_{xx}$ and **b**, $\rho_{xy}$ at $\mu_o H_{\parallel c} = 7$ T for selected currents I; the hollow and solid arrows mark the peak temperature $T_P$ and the Curie temperature $T_C$, respectively. $T_P - T_C \approx 32$ K for all I. **c**, The schematic of the COC Hall effect.

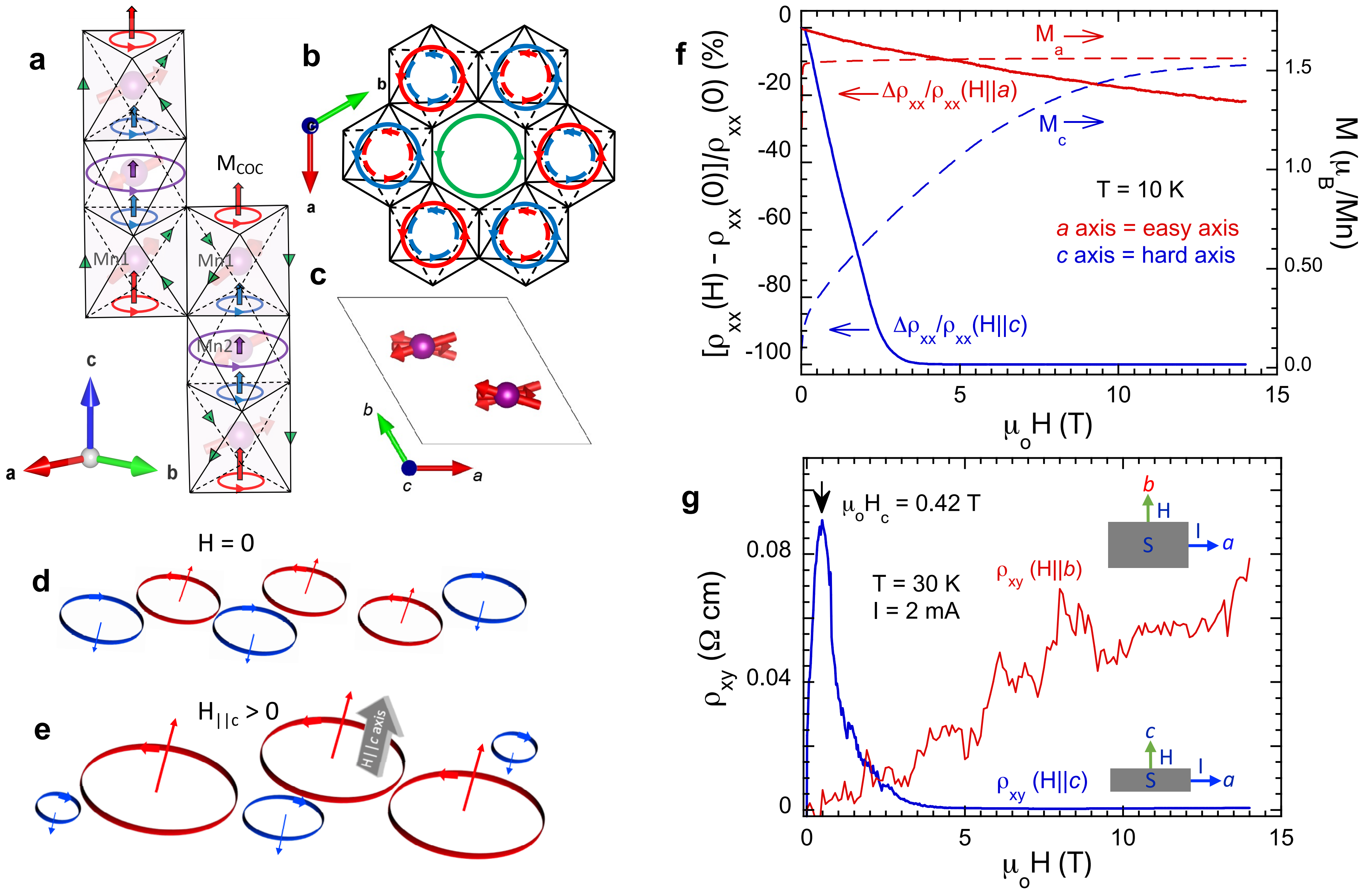

a
b
c
d
e
f
g
M_COC
Mn1
Mn2
H = 0
H||c > 0
H||c axis
M_a
M_c
Δρxx/ρxx(H||a)
Δρxx/ρxx(H||c)
T = 10 K
a axis = easy axis
c axis = hard axis
[ρxx(H) - ρxx(0)]/ρxx(0) (%)
M (μB/Mn)
μoH (T)
μoHc = 0.42 T
T = 30 K
I = 2 mA
ρxy (H||b)
ρxy (H||c)
ρxy (Ω cm)

Figure 1

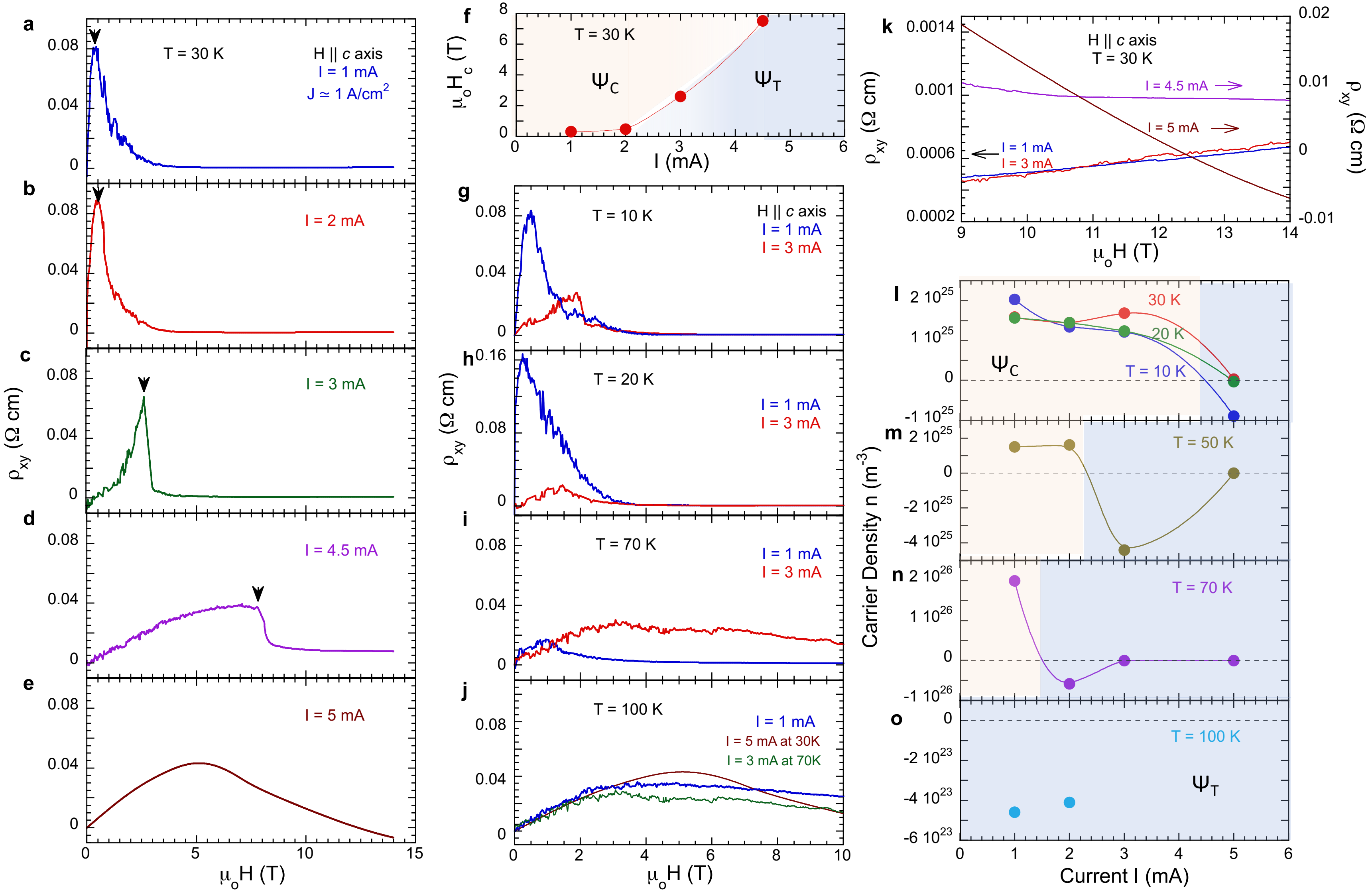
a
b
c
d
e
f
g
h
i
j
k
l
m
n
o
T = 30 K
H || c axis
I = 1 mA
J ≃ 1 A/cm2
I = 2 mA
I = 3 mA
I = 4.5 mA
I = 5 mA
ρxy (Ω cm)
μoH (T)
μoHc (T)
Ψc
ΨT
I (mA)
T = 10 K
T = 20 K
T = 70 K
T = 100 K
I = 5 mA at 30K
I = 3 mA at 70K
30 K
20 K
T = 10 K
T = 50 K
T = 70 K
T = 100 K
Carrier Density n (m-3)
Current I (mA)

Figure 2

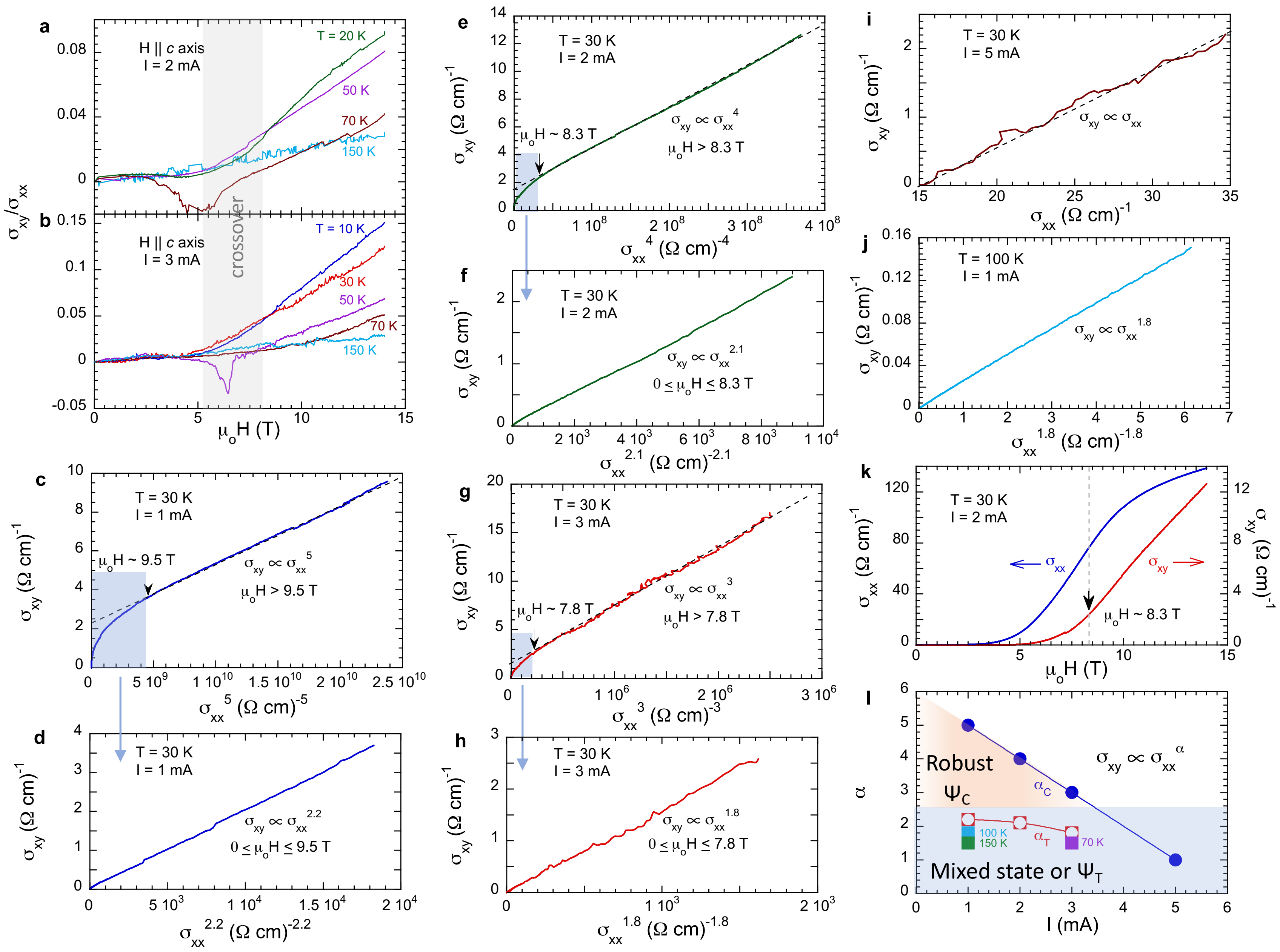
a
H || c axis
I = 2 mA
T = 20 K
50 K
70 K
150 K
b
H || c axis
I = 3 mA
T = 10 K
30 K
50 K
70 K
150 K
crossover
σxy/σxx
μoH (T)
c
T = 30 K
I = 1 mA
μoH ~ 9.5 T
σxy ∝ σxx5
μoH > 9.5 T
d
T = 30 K
I = 1 mA
σxy ∝ σxx2.2
0 ≤ μoH ≤ 9.5 T
e
T = 30 K
I = 2 mA
μoH ~ 8.3 T
σxy ∝ σxx4
μoH > 8.3 T
f
T = 30 K
I = 2 mA
σxy ∝ σxx2.1
0 ≤ μoH ≤ 8.3 T
g
T = 30 K
I = 3 mA
μoH ~ 7.8 T
σxy ∝ σxx3
μoH > 7.8 T
h
T = 30 K
I = 3 mA
σxy ∝ σxx1.8
0 ≤ μoH ≤ 7.8 T
i
T = 30 K
I = 5 mA
σxy ∝ σxx
j
T = 100 K
I = 1 mA
σxy ∝ σxx1.8
k
T = 30 K
I = 2 mA
σxx
σxy
μoH ~ 8.3 T
l
Robust ΨC
αC
αT
σxy ∝ σxxα
100 K
150 K
70 K
Mixed state or ΨT
I (mA)

Figure 3

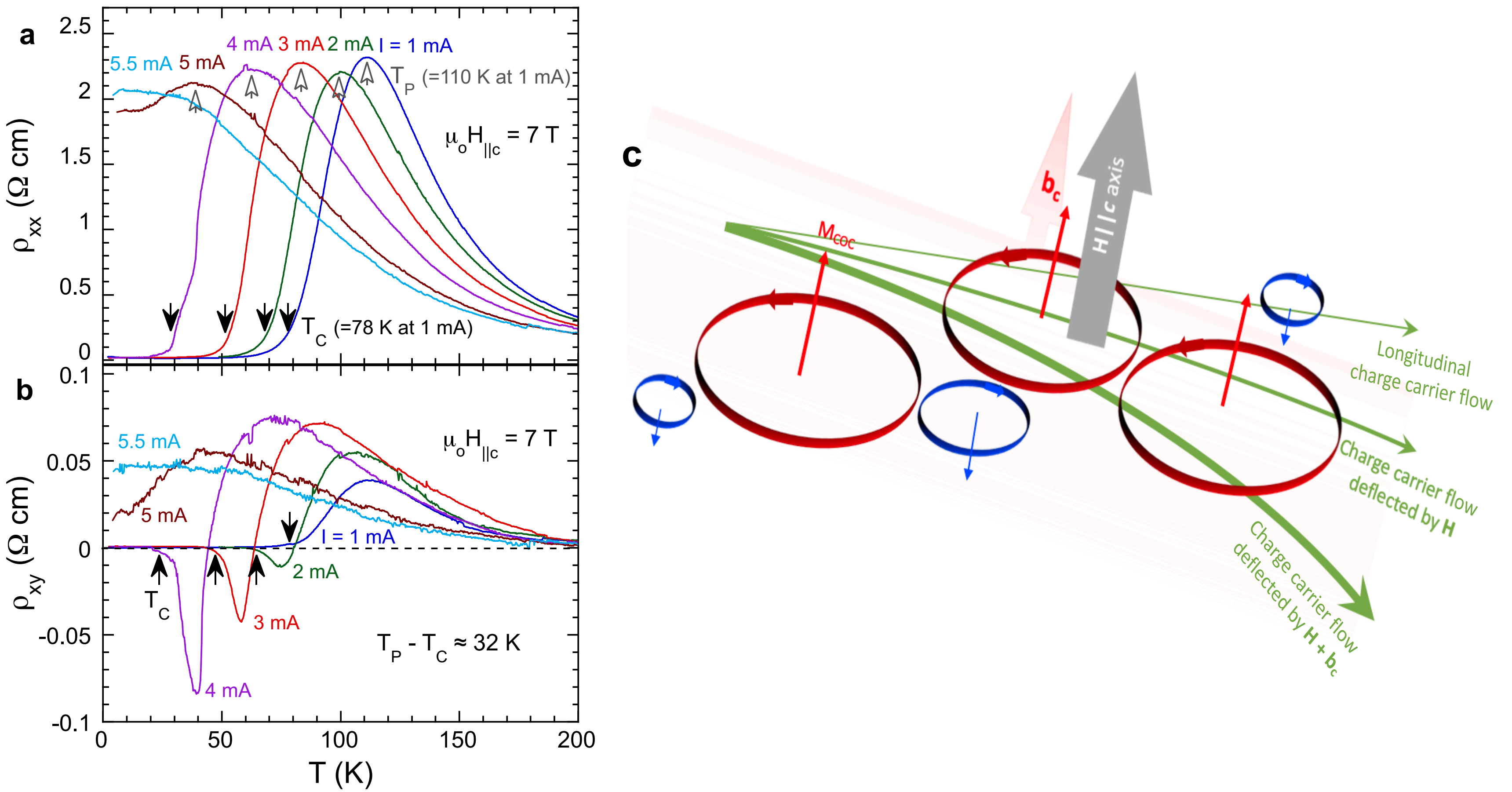

a
b
c
ρxx (Ω cm)
ρxy (Ω cm)
T (K)
5.5 mA
5 mA
4 mA
3 mA
2 mA
I = 1 mA
TP (=110 K at 1 mA)
TC (=78 K at 1 mA)
μoH||c = 7 T
TC
TP - TC ≈ 32 K
MCOC
bc
H||c axis
Longitudinal charge carrier flow
Charge carrier flow deflected by H
Charge carrier flow deflected by H + bc

Figure 4

## Methods

### *Experimental details and data processing*

The Hall resistivity is measured with a Quantum Design DynaCool 14T PPMS. A standard 4-wire configuration for the Hall coefficient measurements is adopted. The applied current is supplied by a Keithley 6220 precision current source, which is paired with a Keithley 2812A nanovoltmeter to measure the Hall voltage using the delta mode.

The measured voltage, $V_{meas}$, is a superposition of the Hall voltage, $V_{xy}$, and the longitudinal voltage, $V_{xx}$, due to the inevitably misaligned voltage leads (similarly the measured resistivity $\rho_{meas}$ is a superposition of $\rho_{xy}$ and $\rho_{xx}$). To eliminate $V_{xx}$, a field H sweep is performed and $V_{xy}$ is determined from the antisymmetrization of $V_{meas}$:

$$V_{xy}(H) = \frac{1}{2}[V_{meas}(H) - V_{meas}(-H)].$$

Similarly, $V_{xx}$ is determined from the symmetrization of $V_{meas}$:

$$V_{xx}(H) = \frac{1}{2}[V_{meas}(H) + V_{meas}(-H)].$$

Because of the nature of the COC domains in $Mn_3Si_2Te_6$, a hysteresis in the Hall measurements is observed between the field ramping up data and the field ramping down data (**Extended Data Fig.1**). To eliminate the effect of hysteresis in the antisymmetrization of $V_{meas}$, we process the data in the following way: The data as a function of H are grouped into 4 parts: 14 T → 0 T (positive H, |H| decreasing), 0 T → -14 T (negative H, |H| increasing), -14 T → 0 T (negative H, |H| decreasing), and 0 T → 14 T (positive H, |H| increasing). The two sets of |H| decreasing data and the two sets of |H| increasing data are regrouped and $V_{xy}$ is obtained from the antisymmetrization of either group. The Hall voltage $V_{xy}$ retrieved this way is proven consistent, independent of the hysteresis.

The Hall voltage $V_{xy}$ is then normalized to the Hall resistivity $\rho_{xy} = V_{xy} \cdot t/I$, where I is the current and t is the thickness of the sample. The longitudinal resistivity, $\rho_{xx}$, is determined in a similar way. The longitudinal conductivity, $\sigma_{xx}$ and the Hall conductivity, $\sigma_{xy}$, are thus given by $\sigma_{xx} = \rho_{xx}/(\rho_{xx}^2 + \rho_{xy}^2)$ and $\sigma_{xy} = \rho_{xy}/(\rho_{xx}^2 + \rho_{xy}^2)$.

***Absence of Joule heating***

Self-heating effects cause a continuous drift in local temperature. They are generally isotropic or diffusive and vary continuously with changing current. Such behavior is ruled out in the present study: The peak in the field dependence of $\rho_{xy}$ occurs and shifts abruptly with increasing external current I (**Figs.2a-2e**). On the other hand, the peak in the field dependence of $\rho_{xy}$ at a constant current remains essentially unshifted with increasing temperature, as shown in **Extended Data Fig. 2**. The contrasting behaviors in **Figs.2a-2e** and **Extended Data Fig. 2** are inconsistent with self-heating effects and rule out thereof. The absence of Joule heating is also confirmed in the data of the field dependence of $\rho_{xy}$ at **H** || ***b*** axis, as shown in **Extended Data Fig. 3**, where the field dependence of $\rho_{xy}$ shows no significant change with increasing I.

***Additional data for scaling relation***

A scaling relation with $\alpha_{LF} \leq 2$ is observed at T = 70 K and I = 3 mA, as shown in **Extended Data Fig.4**. This behavior provides additional evidence that an external current exceeding a critical current $I_C$ can suppresses the COC state below $T_C$ (=78 K).

***Additional notes on the charge carriers***

The charge carriers are primarily holes in $\Psi_C$ and electrons in $\Psi_T$ between $T_C$ and 120 K. Above 120 K, the charge carriers become holes again, as indicated by this and our previous studies [2].

**Acknowledgments**

This work is supported by National Science Foundation via Grant No. DMR 2204811.

**Author contributions**

Y.Z. and Y.F.N. contributed to this work equally. Y.Z conducted measurements of the physical properties and data analysis; Y.F.N. grew the single crystals, characterized the crystal structure of the crystals, and conducted measurements of the physical properties and data analysis. P.S. conducted the theoretical analysis and contributed to paper revisions. R.N. conducted the theoretical analysis and contributed to paper revisions; L.E.D. contributed to the data analysis and paper revisions. G.C. initiated and directed this work, analyzed the data, constructed the figures, and wrote the paper.

**Data availability**

The data that support the findings of this work are available from the corresponding authors upon request.

**Competing interests**

The authors declare no competing interests.

**Materials & correspondence**

Professor Gang Cao, gang.cao@colorado.edu

**Extended data figure legends**

**Extended Data Fig.1. Hysteresis in the measured resistivity $\rho_{meas.}$** A representative set of data for $\rho_{meas}$ at **H || *c*** axis from which $\rho_{xy}$ and $\rho_{xx}$ are obtained. Note the observed hysteresis is consistent with the formation of the COC domains discussed in the main text.

**Extended Data Fig.2. The Hall effect as functions of magnetic field and temperature.** The magnetic field dependence of $\rho_{xy}$ (H||*c*) at I = 1 mA for selected temperatures. Note that the peak $\rho_{xy}$(H) at 1 mA remains essentially unshifted with increasing temperature, in contrast to the rapid shift of the peak with increasing external current I observed in $\rho_{xy}$(H) (**Figs.2a-2e**). The contrasting behaviors help rule out self-heating effects.

**Extended Data Fig. 3. The Hall effect as functions of magnetic field and external current for H || *b* axis.** The magnetic field dependence of $\rho_{xy}$ (H||*b*) at T = 30 K for selected currents. Note the slope of $\rho_{xy}$ (H||*b*) remains essentially unchanged, and there is no sign of AHE even though the *b* (or *a*) axis is the magnetic easy axis, where the magnetization is saturated at $\mu_o H_{||b}$ > 0. 1 T.

**Extended Data Fig. 4. Additional scaling relation.** A scaling relation $\sigma_{xy} \propto \sigma_{xx}^{1.5}$ with $\alpha_{LF}$ = 1.5 is observed at T = 70 K and I = 3 mA. Note that this behavior provides additional evidence that an external current exceeding a critical current can suppresses the COC state below $T_C$.

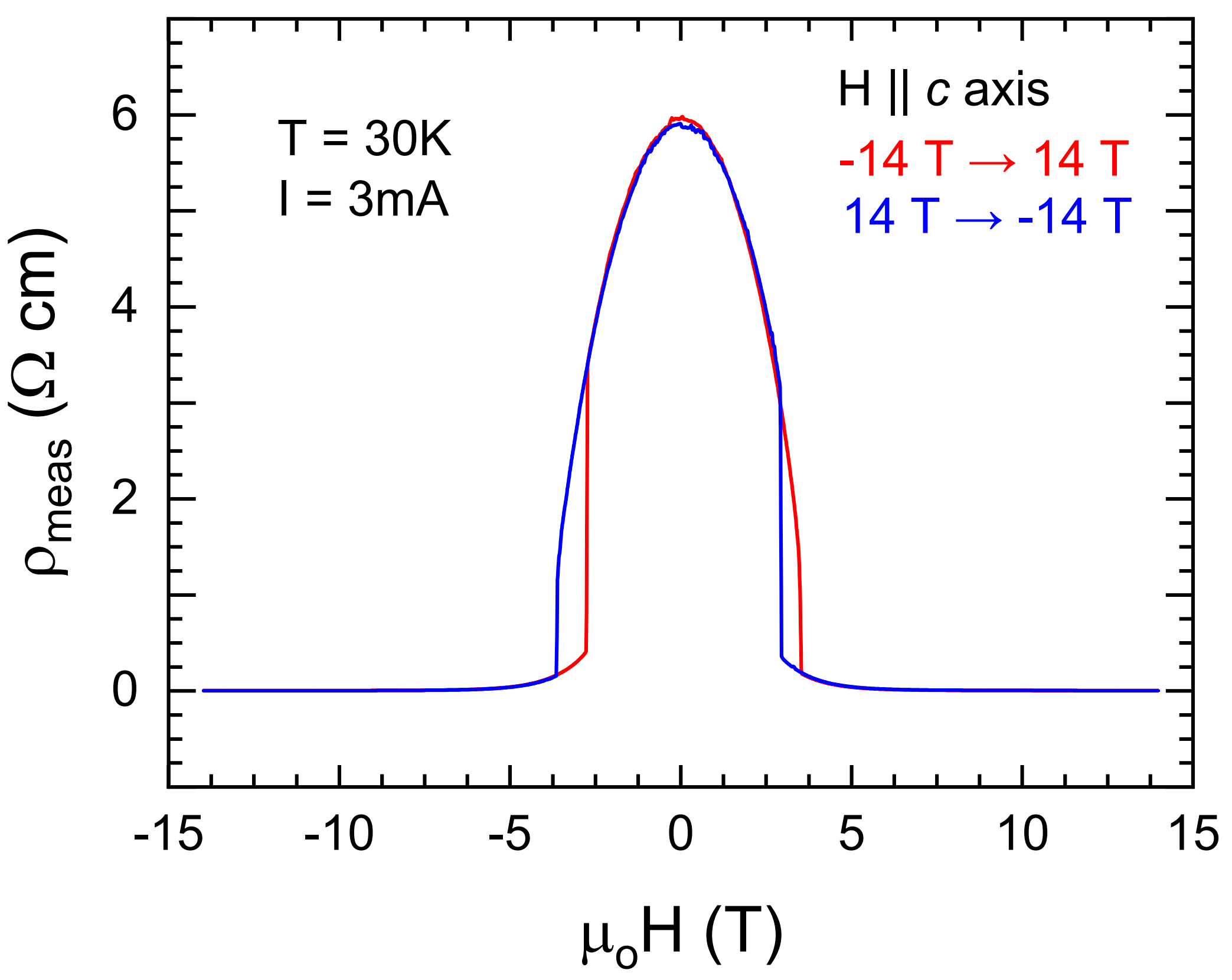

Extended Data Fig. 1

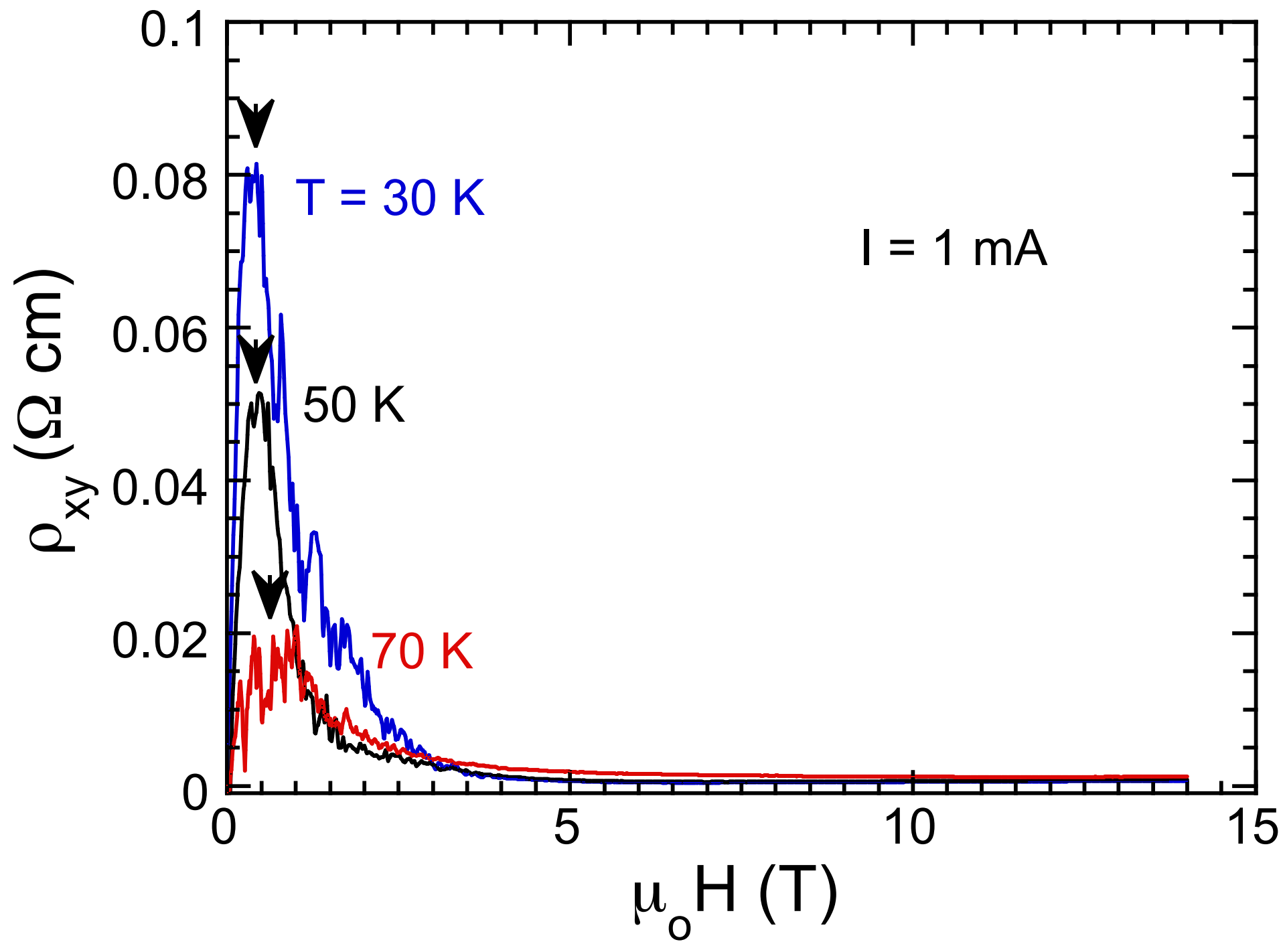

Extended Data Fig. 2

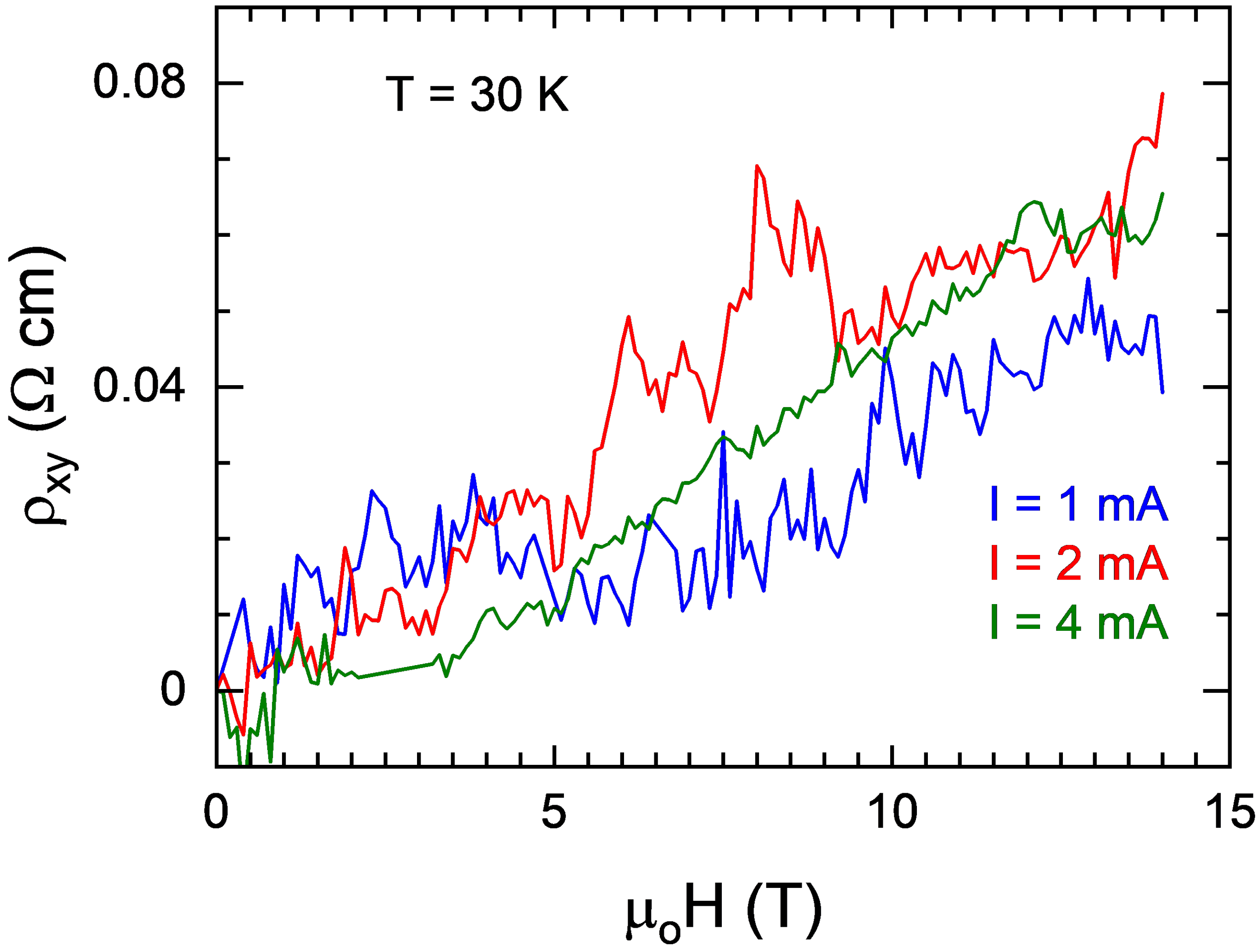

Extended Data Fig. 3

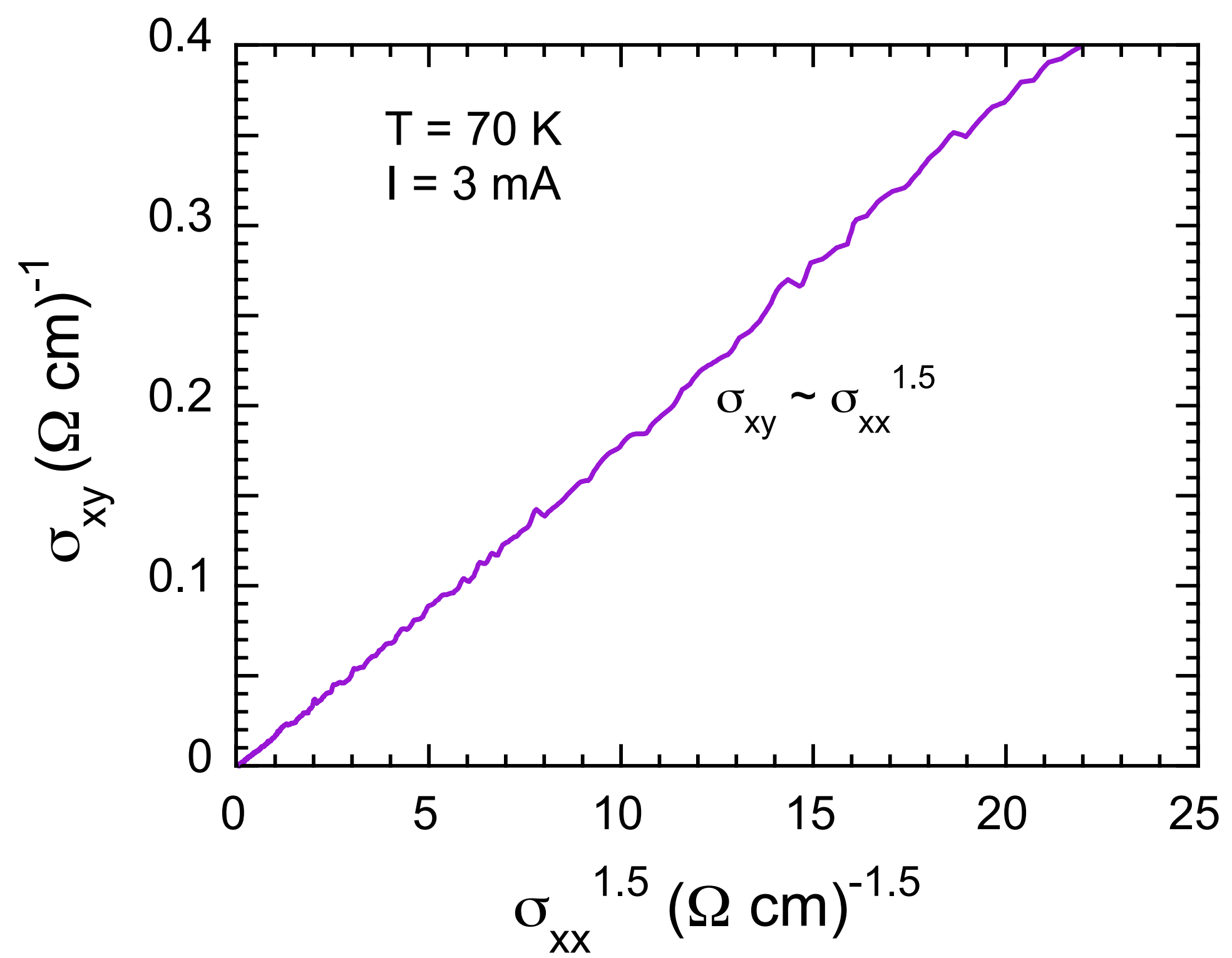

Extended Data Fig. 4